\documentclass[twocolumn,tighten]{aastex631}

\newcommand{\omegacen}{$\omega$~Cen}
\newcommand{\msun}{M_\odot}
\usepackage{amsmath}
\usepackage{enumitem}
\usepackage{soul}

\shorttitle{IMBH Growth in Omega Cen}

\begin{document}

\title{Growing the Intermediate-mass Black Hole in Omega Centauri}

\author[0000-0002-0933-6438]{Elena Gonz\'{a}lez Prieto}
\affil{Center for Interdisciplinary Exploration \& Research in Astrophysics (CIERA) and Department of Physics \& Astronomy, Northwestern University, Evanston, IL 60208, USA}

\author[0000-0003-4175-8881]{Carl L.~Rodriguez}
\affiliation{%
 Department of Physics and Astronomy,
    University of North Carolina at Chapel Hill,
    120 E. Cameron Ave, Chapel Hill, NC, 27599, USA
}%

\author[0000-0002-1270-7666]{Tomás Cabrera}
\affil{McWilliams Center for Cosmology, Carnegie Mellon University,
5000 Forbes Avenue, Pittsburgh, PA 15213, USA}

\begin{abstract}

The recent detection of fast-moving stars in the core of Omega Centauri (\omegacen), the most massive globular cluster (GC) in the Milky Way, has provided strong evidence for the presence of an intermediate-mass black hole (IMBH). As \omegacen\, is likely the accreted nucleus of a dwarf galaxy, this IMBH also represents a unique opportunity to study BH seeding mechanisms and their potential role in the formation of supermassive BHs.  We present Monte Carlo $N$-body models of \omegacen\ with detailed treatments for the loss cone dynamics involving stars, binaries, and compact objects. Starting with BH seeds of $500$--$5000M_{\odot}$ (consistent with runaway collisions of massive stars), our cluster models grow IMBHs with masses of $\sim50{,}000M_{\odot}$ after 12 Gyr, while successfully reproducing the present-day surface brightness and velocity dispersion profiles of \omegacen.  We find a population of fast stars similar to those observed in the core of \omegacen\, with the fastest stars originating from binaries that were tidally disrupted by the IMBH.  The IMBH growth is primarily driven by mergers with $30$--$40 \, M_{\odot}$ BHs, suggesting a present-day IMBH-BH merger rate of $\sim(4$--$8)\times10^{-8}~\rm{yr}^{-1}$ in \omegacen-like GCs.  Our models also predict a similar rate of tidal disruption events ($\sim5\times10^{-8}~\rm{yr}^{-1}$) which, depending on the frequency of \omegacen-like GCs per galaxy, may represent anywhere from $0.1\%$ to $10\%$ of the observed TDE rate.

\end{abstract}
\section{Introduction}

Omega Centauri (hereafter \omegacen) is in many ways a distinct outlier of the Milky Way globular cluster (GC)  population.  In addition to being the most massive GC, it contains a broad metallicity distribution \citep[e.g.,][]{Freeman1975,Frinchaboy2002}, multiple stellar populations \citep[e.g.,][]{Lee1999, Pancino2000}, a peculiar orbit \citep[e.g.,][]{Dinescu1999}, and differences in the spatial and kinematic distributions between metal-poor and metal-rich populations \citep[e.g.,][]{Norris1997, Ferraro2002}. 

Because of these many features, it has been often suggested that \omegacen \, is the remnant nucleus of an accreted dwarf galaxy \citep[e.g.,][]{Tsuchiya2003,Bekki2003,Ideta2004, Limberg2024}. To explore this scenario, \cite{Bekki2003} conducted numerical simulations in which they showed that a merger between a young Milky Way and the \omegacen \, progenitor would result in the outer envelope being stripped away and the remaining nucleus bound in a retrograde orbit, consistent with observations. Furthermore, they find that repeated tidal interactions with the Milky Way can cause periodic radial gas inflow, which can in turn explain the multiple stellar populations observed in \omegacen. 

Due to \omegacen's large mass ($\sim 4 \times 10^6 \msun$) and possible origin, it has been speculated that it harbors an intermediate-mass black hole (IMBH) at its center. In fact, $N$-body models favor an IMBH with a mass in the range $\sim 40{,}000$--$50{,}000 \msun$ \citep{Jalil2012, Baumgardt2017}. However, resolving whether the observed velocity dispersion is a result of a central IMBH or a population of stellar-mass BHs is a challenging task. Notably, \cite{Zocchi2019} showed that a small population of stellar-mass BHs could also produce a rise in the central velocity dispersion. In a later study, \cite{Baumgardt2019b} concluded that the signatures in the velocity dispersion profile were more consistent with a centrally concentrated cluster of stellar-mass BHs than with an IMBH, due to the lack of fast-moving stars. Recently, using more than $20$ years of archival Hubble Space Telescope (HST) data of the inner regions of \omegacen, \cite{Haberle2024} identified $7$ fast-moving stars in the cluster center, providing the most compelling evidence for the presence of an IMBH. Their analysis suggests that the most likely explanation for these sources is an IMBH with a mass between $8{,}200$--$50{,}000 \msun$,  while ruling out alternative explanations such as tight binary systems with stellar-mass BHs or unbound stars. 

An active area of research involving massive BHs (MBHs) in dense stellar systems is the prediction and characterization of tidal disruption events (TDEs). These are observational signatures that arise from the disruption of a star that wanders too close to the MBH \citep{Rees1988}. TDEs are extreme laboratories for probing stellar structures through observed properties of the ejecta \citep{MacLeod2012, Law-Smith2019}, constraining BH mass and spin measurements \citep{Kesden2012, Mockler2019, Huang2024, Mummery2024}, and for studying the loss-cone dynamics around MBHs \citep{Magorrian1999,Wang2004}. The TDE rate has been extensively studied in the context of galactic nuclei, with theoretical estimates ranging from $ \rm 10^{-5} \, yr^{-1} gal^{-1}$ (with an enhancement of up to two orders of magnitude in galaxies with nuclear star clusters (NSCs), see \citet{Pfister2020}) to a $\rm few\times 10^{-4} \, yr^{-1} gal^{-1}$ \citep{Stone2016, Stone2017}. While previous work has focused on supermassive black holes (SMBHs), \cite{Chang2025} recently explored the IMBH regime finding that the IMBH TDE rate increases with BH mass and typically involves deeply plunging events. The upcoming Vera Rubin Observatory Legacy Survey of Space and Time (LSST) is expected to detect thousands of TDEs per year  \citep{Gezari2008A, Strubbe2009, Bricman2020}, offering an unprecedented opportunity to understand the dynamics around IMBHs and discover rare events such as off-nuclear TDEs potentially originating from IMBHs in GCs. 

The presence of an IMBH in a GC is also interesting for its potential to produce intermediate mass ratio inspirals (IMRIs), which will be detectable by future space-based gravitational wave observatories such as the Laser Interferometer Space Antenna (LISA). While extreme mass ratio inspirals (EMRIs) have been widely studied in the context of NSCs where central SMBHs typically have masses $ M > 10^6 \msun$ \citep[e.g.,][]{Hopman2005,Bar-Or2016, Kaur2025}, more recent work has extended these studies to the intermediate-mass regime. Of particular interest is the relative occurrence of true inspirals —where the compact objects undergo many orbits before merger— versus direct plunges \citep[e.g.,][]{Rom2025}. Recent studies \citep[e.g.,][]{Qunbar2024,Mancieri2025} have explored this distinction in the IMBH regime and found that the traditional separation between these classes does not hold. Instead, direct plunges have a non-negligible probability of transitioning into ``cliffhanger'' EMRIs, potentially increasing the EMRI rate significantly. 

The possible origin of GCs as stripped nuclei of dwarf galaxies suggests that their IMBH population could provide insights into the MBH population in low-mass galaxies. Since dwarf galaxies have undergone fewer mergers than their more massive counterparts, they likely conserve most of their initial properties. Thus, they can be studied to understand how the BH mass--galaxy mass relation extends to lower galaxy masses and whether these systems could host the MBH seeds that eventually grew into the SMBHs observed at high redshifts \citep{Volonteri2010, Greene2012, Reines2016}.

In this paper, we present Monte Carlo simulations of \omegacen \, with a central IMBH. In Section~\ref{sec:methods}, we describe the new physical prescriptions implemented to accurately model the loss cone dynamics. We then compare our results to observations of \omegacen \, in Section~\ref{sec:omegacen}. Section~\ref{sec:losscone} discusses IMBH growth, TDEs, and IMRIs. Finally, in Section~\ref{sec:discussion} we address caveats of our method, summarize results, and discuss potential astrophysical implications. 

\section{Methods}
\label{sec:methods}

We perform our simulations using the \texttt{Cluster Monte Carlo} (\texttt{CMC}), a H\'{e}non-type Monte Carlo code that models the evolution of star clusters \citep[see][for the most recent overview]{Rodriguez_2022}. \texttt{CMC} includes prescriptions for key physical processes such as two-body relaxation \citep{Joshi2000}, stellar collisions \citep{Fregeau2007}, and direct integration of small $N$-body strong encounters using \texttt{Fewbody} \citep{Fregeau2004}. Finally, \texttt{COSMIC}, a population synthesis code, is fully integrated into \texttt{CMC} to treat stellar and binary evolution \citep{Breivik19}.

Building on methods developed in previous studies \citep{FreitagBenz2002, Umbreit2012}, we have incorporated new physical prescriptions to accurately model the presence of an IMBH at the center of the cluster. In our approach, the IMBH is treated as a fixed point particle at the cluster's center, acting as an additional gravitational potential. When stars venture too close to the IMBH, they can be tidally disrupted, producing highly energetic transient signals known as TDEs. These disruptions occur on orbital timescales, when the pericenter distance falls within the tidal disruption radius ($R_{\rm tidal}$). Assuming a Keplerian orbit around a non-spinning BH, the disruption radius for a non-rotating star is given by

\begin{equation} \label{eq:Rdisr}
R_{\rm disr} = R_{\rm tidal} \simeq \left( 2 \frac{M_{\rm BH}}{M_\star} \right)^{1/3}R_{\star}
\end{equation}

\noindent where $M_{\rm BH}$ and ${M_\star}$ are the masses of the MBH and star, respectively, and $R_{\star}$ is the stellar radius. For binaries, $R_{\rm disr}$ scales with $a(1+e)$ rather than $R_{\star}$, where $a$ is the semimajor axis of the binary and $e$ its eccentricity. Note that the exact expression for the tidal radius will depend on stellar structure, stellar spin, and BH spin, among others.

The set of velocity vectors that lead to pericenter distances smaller than $R_{\rm disr}$ create a conical region known as the loss cone \cite[see Figure 3 in][]{FreitagBenz2002}. From energy and angular momentum conservation, the loss cone aperture angle can be derived as \nopagebreak

\begin{equation} \label{eq:LC}
\begin{split}
{\sin}^{2}(\theta_{\rm LC}) = 2 \left(\frac{R_{\rm disr}}{vR} \right)^{2} \left[\frac{v^2}{2} + \frac{GM_{\rm BH}}{R_{\rm disr}} \left(1 - \frac{R_{\rm disr}}{R} \right) \right. \\\\
\left. + \Phi_{\star}(R) - \Phi_{\star}(R_{\rm disr})\right]
\end{split}
\end{equation}

\noindent where $v$ is the velocity of the object in the cluster frame, $R$ is the object's position, and $\Phi_{\star}(R) = \Phi(R) + GM_{\rm BH}/R$ is the cluster contribution to the gravitational potential. Note that this expression for the loss cone angle also considers the potential of the cluster at the disruption radius, an important correction for clusters with lower-mass MBHs.

The primary mechanism that replenishes loss cone orbits is two-body relaxation. In \texttt{CMC}, the cumulative effect of many two-body encounters is represented as a deflection angle ($\delta \theta_{\rm step}$) in the encounter frame between neighboring particles \citep[see Figure 1 of ][]{Rodriguez_2022}. While this approach is highly effective for treating local relaxation, it has two serious limitations when applied to clusters with MBHs. First, the effective two-body encounters are applied over a timestep proportional to the local relaxation time. However, in clusters with MBHs, stars and BHs can frequently wander into (and out of) the loss cone on orbital timescales (stars typically undergo thousands to hundreds of thousands of orbits per CMC timestep). Second, the H\'enon method models all dynamics as occurring between neighboring particles. But in a cluster with a MBH, any object can potentially interact with the MBH in a given timestep. Because of these, we require additional prescriptions to accurately model the wandering of objects into the loss cone.

For every star in the cluster, we simulate a random walk of the object's velocity vector during a \texttt{CMC} timestep. Using the radial orbital period, we determine the number of orbits a star will complete within a timestep, and estimate the ``representative'' diffusion angle during a single orbit

\begin{equation} \label{eq:norb}
\delta\theta_{\rm orb} \equiv n_{\rm orb}^{-1/2}\delta\theta_{\rm step}  \, { \, \rm with \, } \, n_{\rm orb} = \frac{\delta t_{\rm step}}{P_{\rm orb}}
\end{equation}

The random walk in the velocity vector is performed such that it covers the same angle as predicted by H\'enon's two-body relaxation scheme for that particle.  See Appendix~\ref{sec:rw_method} for details and comparisons to direct $N$-body simulations. Some orbits will have pericenter distances within the IMBH's Schwarzschild radius ($R_{\rm ss}$), causing the star to be disrupted inside the event horizon, leaving no observational signatures. To account for this, if $R_{\rm ss} > R_{\rm tidal}$, we set $R_{\rm disr}$ equal to $R_{\rm ss}$ and allow for full accretion by the IMBH . Otherwise, we assume $50 \%$ mass loss during a TDE. Binaries within the loss cone are directly integrated as a three-body system (with the IMBH as a tertiary) using \texttt{Fewbody}, allowing us to track their disruption and the dynamical fate of the binary members. Note that, when taking general relativity into account, stars can be captured at radii slightly larger than $R_{\rm ss}$, plunging directly into the MBH without producing observable flares \citep[e.g.,][]{Servin2017}.

To account for the inspiral of compact objects into the IMBH, we calculate the inspiral time at each random walk step using Equations $5.4$ and $5.5$ from \cite{Peters1964}, assuming Keplerian orbits. Note that we are not able to model the change of orbits as they inspiral into the IMBH. This would require us to simultaneously recalculate the potential after each random walk step for all objects, which is computationally intractable. Instead, we assume that if an object has a short enough inspiral timescale, it will inspiral in an isolated manner. After the random walk ends, $dE/dT$ and $dJ/dT$ due to gravitational wave (GW) emission are calculated for all objects, and applied to their orbits during the next \texttt{CMC} timestep.

\subsection{Initial Conditions}

Out of a grid of $35$ runs, we present our best-fit models with an initial number of $N = 1.1 \times 10^7$ particles, IMBH seeds of $500 \msun$ and $5000\msun$, virial radius of $5$~pc, and a tidal radius of $208.6$~pc \citep{Baumgardt2017}. The initial stellar distribution follows an Elson profile with $\gamma = 3$ \citep{Elson1987} and a bottom-heavy Kroupa initial mass function (IMF) with $\alpha_3 = 2.5$ \citep{Kroupa2001}. While the Elson profile is typically used for modeling young star clusters, we note that at large radii, the mass density Elson profiles with $\gamma = 3$ go as $\propto r^{-1/4}$, which is the same as Jaﬀe and Hernquist models \citep{Jaffe1983, Hernquist1990}, commonly used to describe spherical galaxies and bulges, which is particularly relevant in the case of \omegacen. Finally, we assume a stellar metallicity of [Fe/H]$=-1.7$, an initial $2\% $ binary fraction \citep{Wragg2024}, and a cluster age of $12 \, \rm Gyr $ \citep[approximate mean stellar age of Omega Cen, see][]{Clontz2024}. For the primordial binaries, separations are sampled from a uniform distribution in logarithmic space \citep{Abt1983} (up to the local hard-soft boundary for each binary), and eccentricities follow a thermal distribution \citep{Heggie1975}. While only two models are shown, we explored a wide range of initial conditions -- varying the initial numbers of objects ($10^7$ to $1.2 \times 10^7$), virial radii ($3.5$ to $8$ pc), density profiles (Elson profiles with $\gamma = 3$--$6$) and initial mass functions ($\alpha_3 = 2.3$--$2.7$).

Over the cluster lifetime, the $500 \, \msun $ seed grows to \mbox{$\sim 47,000 \, \msun$}, while the $5000 \, \msun $ seed reaches \mbox{$\sim 51,000 \, \msun$}.  In Section~\ref{sec:losscone}  we will examine the growth mechanisms in more detail. Notably, only models with bottom-heavy IMFs prevent excessive growth of the IMBH seeds and remain consistent with observational data. This suggests that the IMBH may be a product of a collisional runaway, in which most of the massive stars in the cluster have merged \citep[e.g.,][]{Gurkan2004, GonzalezPrieto2024,Sharma2025}.

\section{Comparison to Omega Cen Observations}
\label{sec:omegacen}

The likely origin of \omegacen \, as the accreted nucleus of a dwarf galaxy adds significant complexity to its dynamical modeling. Its accretion implies a history of close tidal interactions with the Milky Way, which are expected to have strongly shaped its outer layers. Although some studies have attempted to reconstruct the dynamical history of \omegacen \, \citep{Tsuchiya2003, Bekki2003}, uncertainties about the physical and dynamical properties of the initial dwarf galaxy remain. Additionally, \omegacen \, has a large spread in metallicity, indicating multiple star formation epochs \citep[e.g.,][]{Freeman1975, Johnson2010, Nitschai2024}, as well as plane-of-sky rotation \citep[e.g.,][]{vanLeeuwen2000, Bianchini2018, Haberle2024b}. These characteristics make it challenging to model \omegacen \, with codes such as \texttt{CMC}, which are restricted to spherical symmetry and a single episode of star formation (the latter limitation being shared by all current direct $N$-body and Monte Carlo codes). Despite this, we find remarkable agreement with observed properties, such as surface brightness and velocity dispersion profiles (SBP and VDP, respectively). In other words, while our model may over-simplify the earliest evolution of \omegacen, its excellent agreement with the present-day observation suggest it to be highly reliable over the last $\sim 10$~Gyr, roughly corresponding to the time of the last significant burst of star formation \citep[e.g.,][]{Clontz2024}.

\subsection{Surface Brightness and Velocity Dispersion Profiles }
\label{sec:SBP_VDP}

\begin{figure*}[t]
    \centering
    \includegraphics[width=\linewidth]{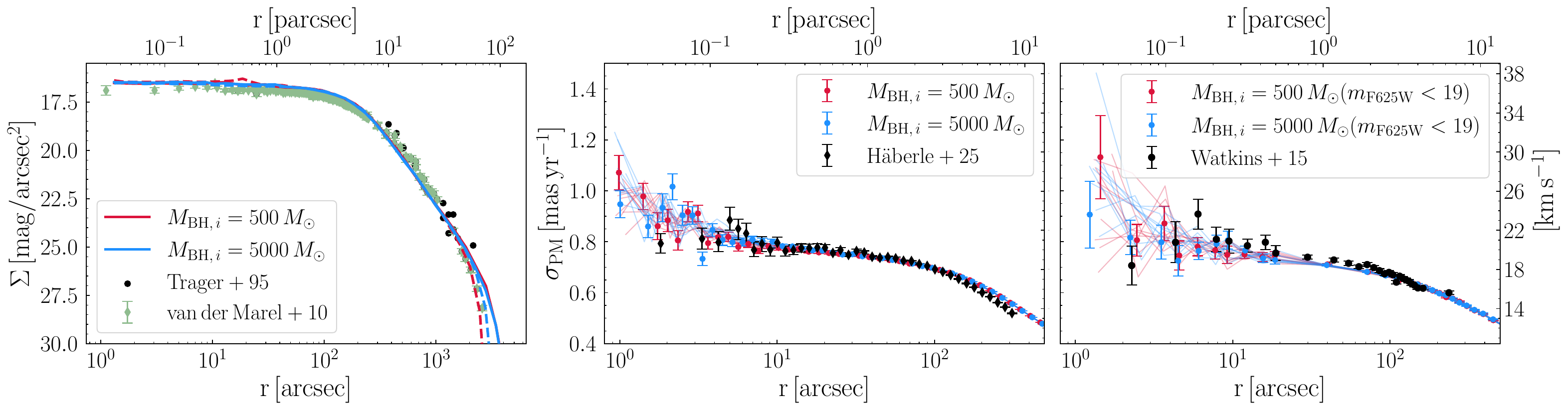}
    \caption{\label{fig:VDP_SBP}\textit{Left panel}: V-band surface brightness profile at $12$~Gyr. Only stellar objects with masses above $0.559 \, \msun$ are included, corresponding to B-band instrumental magnitudes brighter than $-10$, consistent with the quality cuts applied in \citet{Marel2010}. Their data is shown in green diamonds. Round symbols in black are a compilation of ground-based observations from \cite{Trager1995}. An extinction factor of $\rm A_v = 0.372$ is applied to our data \citep{Harris1996}. The dashed lines are models that were evolved with a time-varying tidal tensor representative of \omegacen's orbit for the last Gyr. \textit{Middle panel}: Proper motion velocity dispersion profile at $12 \,$~Gyr shown as points. Lightly shaded lines correspond to profiles between $11.9$ and $12.1 \, \rm Gyr$. Only stellar objects with \mbox{$\rm 13.9 < m_{F625W} < 24 \,$} are included, consistent with the magnitude cuts in \cite{Haberle2024}. The errors are computed as $\sqrt{\sigma^2/ (2N)}.$ \textit{Right panel}: we further limit the stellar population to stars with $m_{\rm F625W} < 19$, and follow a similar binning procedure as detailed in \cite{Watkins2015}, with the caveat that we exclude stars in the inner arcsecond.}
\end{figure*}

The SBPs are computed by assuming a distance of $5.494$ kpc to \omegacen \, \citep{Haberle2025}, excluding compact objects and stars with masses below $0.559\,  \msun$, corresponding to a B-band instrumental magnitude $ < -10 $. This is consistent with the quality cuts applied in HST observations \citep{Marel2010, Anderson2010}. We have also applied an extinction factor of $\rm A_v = 0.372$ \citep{Harris1996, McLaughlin2005}. The V-band SBPs from our simulations are shown in the left panel of Figure~\ref{fig:VDP_SBP}, alongside observed profiles.

The models reproduce the observed data quite closely, only predicting a slightly brighter inner core. Note that there are significant differences in the methods used to compute the SBP. For instance, \cite{Marel2010} use the projected number densities to derive a surface brightness by assuming a distance of $4.8$~kpc. In contrast, our method computes apparent magnitudes directly from the intrinsic luminosity of stars, thereby avoiding the need to infer surface brightness from stellar densities.

Furthermore, we have assumed a fixed tidal field throughout the evolution of the cluster. That is, however, not true about \omegacen \,, which is in an eccentric orbit with a pericenter distance of $1$ kpc and apocenter of $6$ kpc \citep{Dinescu1999}. To account for this, we restart our model at $11$ Gyr and apply a time-varying Galactic tidal field representative of \omegacen's orbit. The tidal field was generated by integrating the present-day position and velocity for \omegacen \, from \cite{Baumgardt2019a} backward in the \texttt{MilkyWayPotential2022} from the \texttt{Gala} galactic dynamics package \citep{gala}.  The spatial second derivatives of the potential (the tidal tensor) were then calculated along that orbit, and used to compute the instantaneous tidal boundary of the cluster over time \cite[see][for details]{Rodriguez2023}.  The resulting profiles at $12$~Gyr appear as dashed lines in the left panel of Figure~\ref{fig:VDP_SBP}. As shown, the close encounter with the Milky Way effectively removes stars near the tidal boundary, resulting in improved agreement with the observed outer layers of the cluster. 

For the proper motion velocity dispersion profiles in the middle panel of Figure~\ref{fig:VDP_SBP}, we include only stars with \mbox{$\rm 13.9 < m_{F625W} < 24 \,$}, require at least $100$ stars per bin, and treat the radial and tangential proper motion measurements as independent measurements, consistent with the procedure followed in \cite{Haberle2025}. The magnitude limits correspond to stellar masses between $0.2$--$0.8 \msun$. The lightly shaded lines correspond to profiles between $11.9$ and $12.1 \rm \,  Gyr$, highlighting the variability in time, whereas the points represent the profile at exactly $12 \rm \, Gyr$. Our models show strong agreement with the observed VDPs, including the inner rise in velocity dispersion, a likely signature of an IMBH at the cluster's center. The high uncertainties in the inner arcseconds reflect the low number of stars in those radial bins. We found that applying a time-varying tidal field does not have any significant effects on the VDP profile, thus we do not include that model in the figure. 

In the right panel of Figure~\ref{fig:VDP_SBP} we compare to the proper motion velocity dispersion measurements of \cite{Watkins2015}. To make a fair comparison, we further limit our sample to only the bright stars ($m_{\rm F625W} < 19$) and follow the binning procedure outlined in that work (with $N_{\rm star} = 25$) with one exception: we exclude stars within the inner arcsecond of the cluster. This ensures a more consistent comparison to the observed data, which does not have values within the inner $\sim 2.5$ arcseconds. Furthermore, this excludes very tightly bound stars originating from binary disruptions, which are expected to get tidally disrupted.

Recently, \cite{Haberle2025} published line-of-sight (LOS) rotation profiles for \omegacen, which could impact the observed velocity dispersion profiles. While we can't model cluster rotation self-consistently in \texttt{CMC}, we tested its impact on the velocity dispersion profiles by incorporating it post-processing. We did this by adding the LOS rotation profile (see their Figure 6) to the LOS velocity from our models, taking into account the cluster rotation axis. We found that it did not significantly alter our VDPs. 

\subsection{Fast-moving stars}
\label{sec:fast_moving}

A key signature of the potential presence of an IMBH at the cluster center is the detection of stars moving faster than the inferred escape speed from the cluster. Such observations invoke a ``missing'' mass component to explain fast-moving objects that are still bound to the cluster. The fast stars reported in \cite{Haberle2024} were compared against an inferred cluster escape velocity of $62$ km/s. 

A crucial mechanism that might bring stars into close orbits around the IMBH is the Hills mechanism, where one of the stars becomes tightly bound to the IMBH while the other is ejected as a hyper-velocity star \citep{Hills1988,Ginsburg2006}. This process has been invoked to explain the S-star population around the galactic center and it yields two key observables. First, it places stars on tight orbits around the IMBH, where they are likely to be tidally disrupted. Second, it ejects the companion at high velocity, some of which have been observed as hyper-velocity stars \citep[e.g.,][]{Brown2005, Boubert2018,Koposov2020}.

\begin{figure}
    \centering
    \includegraphics[width=1\linewidth]{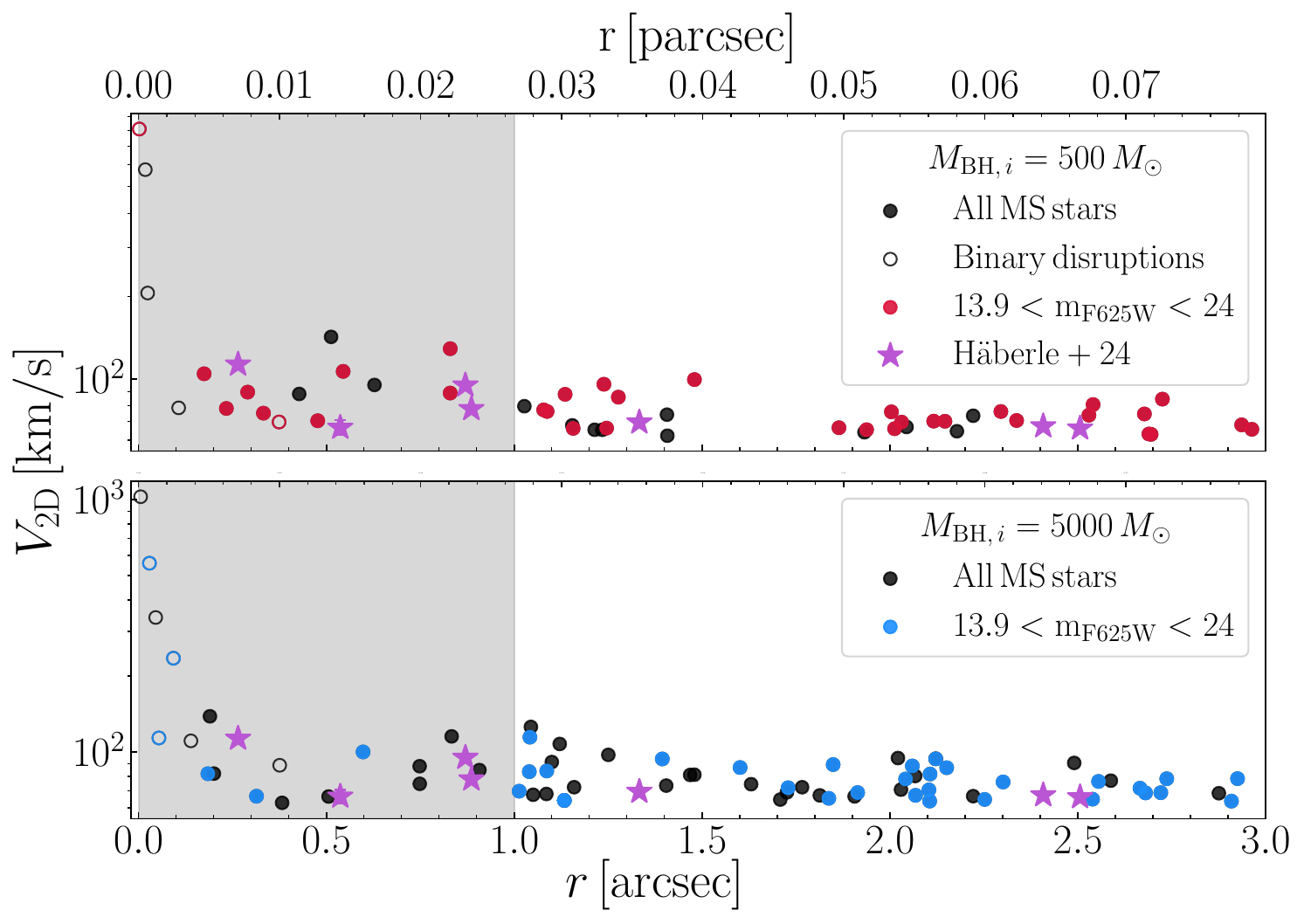}
    \caption{\label{fig:fast_stars} Main sequence stars with projected $2$D speeds larger that $62$ km/s. The initial sample (shown in black circles) is filtered to include only stars with \mbox{$\rm 13.9 < m_{F625W} < 24 \,$} (shown in colored circles, consistent with the observed stellar population in \cite{Haberle2024}, shown as purple stars). The empty circles indicate stars that are remnants of a binary disruption. The shaded gray region shows the $1''$ uncertainty in the cluster center. }
\end{figure}

Figure~\ref{fig:fast_stars} shows the fast-moving MS stars from our models in black, with colored points indicating those that meet the magnitude cutoffs used in \cite{Haberle2024}. Empty circles indicate stars produced by binary disruptions during interactions with the IMBH. Our models predict a larger number of fast-moving stars than has been detected in HST observations. It is important to note that observational data were subject to additional selection criteria (such as measurement quality or the number of detections per source), which we are not able to account for in our simulations. Thus, we expect our population of fast stars to be higher than the detected number. 

The centrally concentrated population of fast stars within the inner $0.2$~arcseconds mainly originates from binary disruptions. They are tidally disrupted within a few million years in most cases (see Section~\ref{sec:TDEs}) and may not have been detectable in HST analyses due to their extremely tight orbits. The number of fast-moving stars from binary disruptions may be sensitive to the assumed properties of the initial binary population. As for the previous binary companions of these bound stars, each of our models produce roughly $1700$ runaway stars, including $46$ ejected within the last $500$~Myr. These stars have a mean escape speed of $206$~km/s, with some reaching velocities as high as $2000$~km/s. The tightly bound stars may be resolved in future JWST observations of \omegacen, while their high-velocity companions could be detected with Gaia.

\section{Massive Black Hole Growth} 
\label{sec:losscone}

In gas-free environments, MBH seeds grow primarily through two main mechanisms: tidal disruption of stars and the inspiral of BHs. The degree of growth depends on both the mass of the initial seed, the degree of mass segregation, and the efficiency of the environment at bringing objects into the loss cone. 

\begin{figure}
    \centering
    \includegraphics[width=\linewidth]{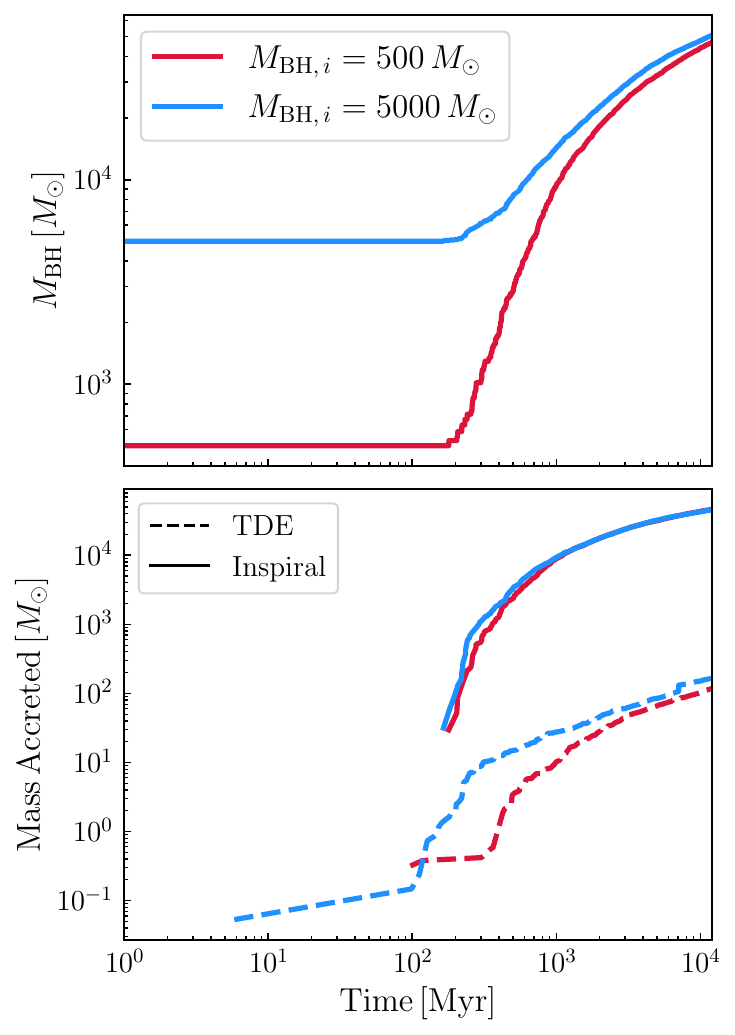}
    \caption{\label{fig:BH_growth} \textit{Upper panel}: mass of the BH seed as a function of time. \textit{Lower panel}: cumulative mass accreted from tidal disruption events (inspirals) in dashed (solid) lines.}
\end{figure}

Over the cluster's $12 \rm \, Gyr $ lifetime, the $500 \, \msun$ seed grows by a factor of approximately $94$, while the $5000 \, \msun$  seed grows in mass by roughly an order of magnitude. Both seeds reach final IMBH masses in the range $47{,}000$--$51{,}000 \msun$, shown in the upper panel of Figure~\ref{fig:BH_growth}. The lower panel shows that the vast majority of this growth is from compact object inspirals. On average, the mass accreted from TDEs is $\sim 143 \, \msun$.

\subsection{Tidal Disruption Events}
\label{sec:TDEs}

There are generally two regimes in which tidal disruption occurs. When orbital timescales ($T_{\rm orb}$) are much longer than the local relaxation timescales ($T_{\rm relax}$), the loss cone is efficiently refilled with stars, and we refer to this as the  ``full loss cone''. Conversely, if $T_{\rm orb} \ll T_{\rm relax}$, this is known as the ``empty loss cone'' regime since stars are removed on orbital timescales and the loss cone is not replenished until after a relaxation timescale. 

A key property of TDEs is the penetration parameter $\beta$, which is the ratio of the tidal radius to the pericenter distance. For values of $\beta \sim 1$, the TDE is a mildly grazing encounter, while TDEs with $\beta >1 $ are deep plunges. In the empty loss cone, objects will undergo many orbits before being tidally disrupted, yielding a distribution of $\beta$ values peaked at $\beta \sim 1$ \citep[e.g.,][]{Lightman1977, Stone2016}. On the other hand, orbits in the full loss cone (or ``pinhole'') regime can enter the loss cone within one orbital period, so the chances of high penetration parameters increase, resulting in a more extended $\beta$ distribution that can be represented by $dN/d\beta \propto \beta^{-2}$.  Deep plunging events could be particularly interesting in the case of white dwarfs, as they may lead to strong compression of the star,  potentially producing GW emission \citep{Stone2013} or releasing nuclear energy comparable to Type Ia supernovae \citep{Rosswog2009}.

\begin{figure}

    \centering
    \includegraphics[width=\linewidth]{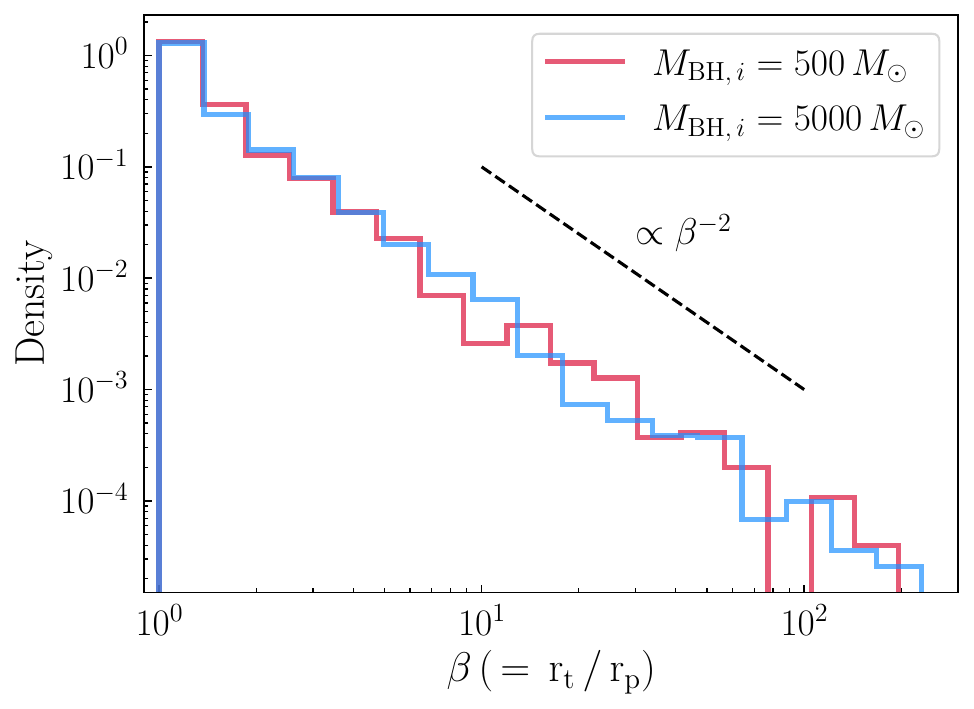}
    \caption{\label{fig:beta} Distribution of penetration parameters ($\beta$) for TDEs. Here, we consider only TDE events involving single stars where the disruption occurred outside of the Schwarzschild radius. In a dashed line we show the theoretical prediction from the full loss cone derived in \citep[e.g.,][]{Lightman1977, Stone2016}.}
    
\end{figure}

In Figure~\ref{fig:beta}, we show the distribution of $\beta$ for the TDEs. The sharp peak at \mbox{$\beta \sim 1$} suggests that many of the TDEs likely originate from the empty loss-cone regime. Indeed, on average, $54\% $ of the TDEs have \mbox{$\beta < 1.5$} while the remaining $46\% $ likely originate from the full loss cone. We also see that the tail of the distribution closely follows the expected \mbox{$dN/d\beta \propto \beta^{-2}$} relation. 

\begin{figure}

    \centering
    \includegraphics[width=\linewidth]{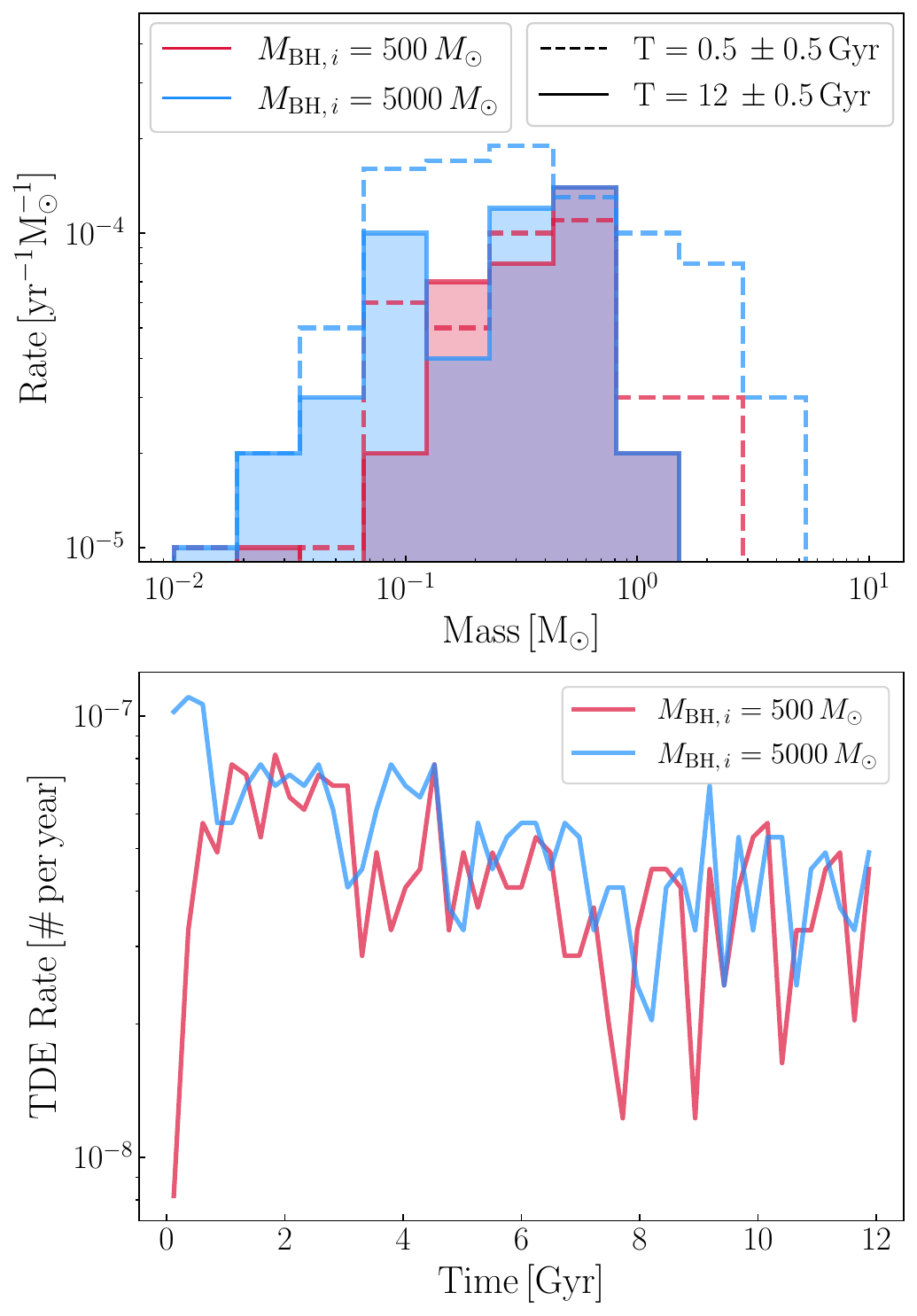}
    \caption{\label{fig:tde} \textit{Top panel}: Distribution of stellar masses involved in tidal disruption events for two time intervals, shown at $0.5 \pm 0.5 \rm \, Gyr$ (dashed lines) and $12 \pm 0.5 \rm \, Gyr$ (shaded). 
    \textit{Bottom panel}: TDE rate as a function of time. Tidal disruptions of both components of a binary are counted as separate events.}
    
\end{figure}

Because \omegacen \, is an old stellar population, the majority of our TDEs originate from low-mass MS stars, see the top panel of Figure~\ref{fig:tde}. At $0.5$~Gyr, the typical TDE mass is $0.55 \, \msun$, decreasing to $0.38 \, \msun$ at $12$~Gyr. Additionally, the total number of TDEs decreases with time due to the gradual expansion of the cluster, causing lower densities near the IMBH. 

In both models, approximately $55\%$ of TDEs involving a single star originated from stars that were initially members of binary systems disrupted during interactions with the IMBH. These stars were subsequently placed in highly eccentric orbits about the MBH (mean eccentricity of $0.97$) and were eventually disrupted. To investigate the impact of the binary fraction on the TDE rate, we ran a model with $10\% $ primordial binary fraction. In that model, $88\%$ of the single-star TDEs originated from stars that had been left over in the vicinity of the IMBH following a binary disruption. This led to an increase in the TDE rate by approximately an order of magnitude, highlighting the important role the binary fraction may play in the observed number of TDE events in all dense stellar environments. 

In the model with the $5000 \msun$ seed, the TDE rate peaks at $\rm \sim 10^{-7} \, yr^{-1}$ at early times and decreases to $\rm 5 \times 10^{-8} \, yr^{-1}$ in the local universe, shown in the bottom panel of Figure~\ref{fig:tde}. In contrast, the lower-mass IMBH model begins with a rate that is an order of magnitude lower but converges to a similar rate at late times. The TDE rate in the first Gyr is lower in the model with the less massive seed because the size of the loss cone scales with the mass of the IMBH.

To estimate a per-galaxy rate, we assume a lower limit of one \omegacen-like cluster per MW-like galaxy and an upper limit of $25$, based on the known population of $170$ GCs in the MW \citep{Vasiliev2021} and the estimate that $15$\% may have originated as dwarf galaxy nuclei \citep[see][]{Kruijssen2012} \footnote{The number of \omegacen-like systems hosting a MBH in MW-like galaxies may be higher, as suggested by the high MBH occupation fraction observed in ultra-compact dwarf galaxies and massive GCs \citep{Voggel2019}.}. This yields a TDE rate in the range $\rm 5 \times 10^{-8} - 1\times10^{-6} \, yr^{-1} gal^{-1}$. 

The observed TDE rate inferred from detections at different bands is \mbox{$ \rm few \times  10^{-5} \rm yr^{-1} \, gal^{-1}$} \citep{Donley2002, Gezari2008, vanVelzen2016, Yao2023}. For SMBHs in galactic nuclei, \cite{Stone2016} estimated a per-galaxy TDE rate of a \mbox{$ \rm few \times 10^{-4} \, yr^{-1} \, gal^{-1}$}, while \cite{Pfister2020} found a rate of $ \rm 10^{-5} \, yr^{-1} \, gal^{-1}$. TDE rates for IMBHs have been estimated in the ranges \mbox{$\sim 10^{-8}$--$10^{-4} \rm \, yr^{-1} \, gal^{-1}$} \citep{Chang2025} and  \mbox{$\sim 10^{-7}$--$10^{-5} \rm \, yr^{-1} \, gal^{-1}$} \citep{Hannah2025}. For those originating from GCs in particular, \cite{Tang2024} estimate a TDE rate of \mbox{$\rm 10^{-8}$--$ 10^{-5} \rm yr^{-1}$} (depending on the cluster' core density and age). 

Although our TDE rate estimates fall at the lower end of these predicted values, our models suggest that $1/1000$--$1/10$ observed TDEs may be off-nuclear and originating from IMBHs in GCs, an exciting prospect as the number of detected TDEs is expected to increase by several orders of magnitude with the upcoming Vera C. Rubin Observatory. Furthermore, previous studies have shown that dense star clusters with top-heavy IMFs and high binary fractions produce IMBHs \citep{Gonzlez2021, DiCarlo2021, GonzalezPrieto2024, Sharma2025}, further increasing the fraction of IMBH-hosting GCs and TDE rate. 

\subsection{BH Captures}
\label{sec:IMRIs}

\begin{figure}

    \centering
    \includegraphics[width=\linewidth]{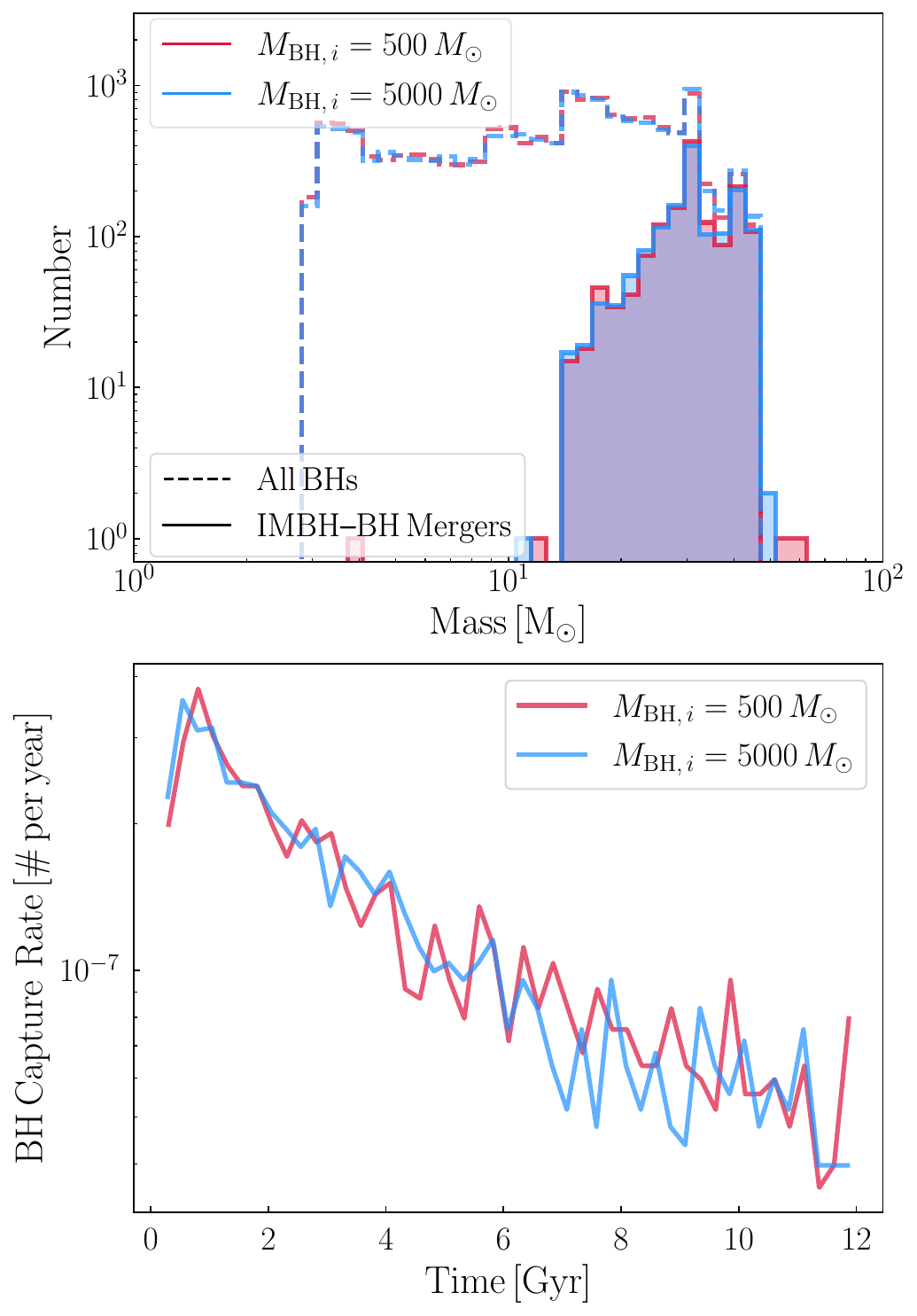}
    \caption{\label{fig:imri} \textit{Top panel}: Distribution of BH masses formed from stellar collapse (dashed) and masses of inspiraling BHs (shaded). \textit{Bottom panel}: BH capture rate as a function of time.  }
\end{figure}

As stated above, the vast majority of BH growth in our \omegacen\, models is driven by mergers with stellar-mass BHs. The top panel of Figure~\ref{fig:imri} shows the distribution of BH masses that eventually inspiral into the IMBH over the cluster's lifetime. As expected from mass segregation, the BHs form a steep density cusp in the core (see Appendix~\ref{sec:cusps}) and the most massive BHs dominate the inspiral events. In fact, the mean mass of inspiraling BHs is $31 \, \msun$, which is substantially higher than the commonly assumed value of $10 \, \msun$ in semi-analytic studies \cite[e.g.,][]{Hopman2006, Rom2024}. This highlights the crucial role that an extended BH mass spectrum and mass segregation play in the growth of IMBHs in dense stellar environments. Furthermore, the extent of this growth is sensitive to the assumed initial-final mass relations and BH natal kick prescriptions. Future studies will explore the impact of different BH recipes on IMBH growth.

Another important prediction from our models is the IMRI rate. For the galactic center, \cite{Bar-Or2016} estimate an EMRI rate of $\rm (1$--$\rm 3) \times 10^{-6} \, yr^{-1}. $ \cite{Rom2024} estimate an EMRI rate of $0.2$--$2.6 \times 10^{-7} \rm \, yr^{-1}$ for a $\rm 4\times10^6 \msun$ MBH, depending on the number fraction of stellar BHs. \cite{Qunbar2024} recently showed that for low-mass IMBHs, there is a $\mathcal{O}(1)$ probability that plunges transition into EMRIs, increasing the EMRI rate by an order of magnitude. 

In our models, we estimate a local BH capture rate in the range $ \rm  (4$--$8) \times 10^{-8} \, \rm yr^{-1}$, shown in the bottom panel of Figure~\ref{fig:imri}. This is comparable to the lower end of the predicted EMRI rates for the galactic center, suggesting that IMBHs in GCs might contribute significantly to the observed IMRI populations. 

However, it should be noted that \texttt{CMC} does not resolve the changes of orbits due to GW emission at each pericenter passage. As a result, our BH capture rates should be interpreted as rough estimates. Furthermore, we do not distinguish between true IMRIs and direct plunges, as this would require self-consistently resolving the inspiral within the code. 

\section{Discussion \& Conclusions}
\label{sec:discussion}

Expanding on previous methods, we have developed an improved approach to accurately model the presence of an IMBH in a star cluster, including the loss cone dynamics. To the best of our knowledge, \texttt{CMC} is the only Monte Carlo code capable of simultaneously modeling two-body relaxation, loss cone dynamics, strong dynamical encounters, and stellar evolution, while remaining computationally inexpensive. 

\begin{itemize}
  \item We have presented our best-fit models for \omegacen \, with initial seed masses of $500 \, \msun$ and $5000 \, \msun$, showing that the seeds grow to $47{,}000$ and $51{,}000 \, \msun$, respectively. Both of our best-fit models have bottom-heavy IMFs, consistent with seed formation from runaway stellar collisions. 
  
  \item We estimate a local TDE rate originating from \omegacen-like clusters in the range \mbox{$\rm 4 \times 10^{-8}$ to $\rm 1\times10^{-6} \, yr^{-1} gal^{-1}$}, which is roughly $1/1000$--$1/10$ of the observed rate. We also find that a higher number of primordial binaries increases the TDE rate by at least an order of magnitude. 

  \item We predict a small population of fast-moving stars produced by binary disruptions within the innermost $0.2$~arcseconds. Notably, these tightly bound stars were responsible for $55\%$ of the single-star TDEs in our models.

  \item We find a BH capture rate in the range \mbox{$ \rm  (4$--$8) \times 10^{-8} \, \rm yr^{-1}$}, which is comparable to the EMRI rates for the galactic center. However, we do not resolve whether these sources will be true inspirals or direct plunges. Nevertheless, this suggests that IMBHs in dense star clusters may significantly contribute to future IMRI detections.
\end{itemize}

To reduce computational costs, we approximate orbits as Keplerian when calculating the inspiral time for all objects in the cluster. This is a reasonable assumption for orbits within the IMBH's influence radius, but starts to break down at larger distances. Additionally, we do not take into account the BH spin or changes in the orbit due to GW emission, which would have implications for the loss cone dynamics and the IMRI signatures. For example, it has been shown that spin will affect the ratio between plunges and EMRIs \citep{Qunbar2024} for IMBHs. We also neglect the possibility that the IMBH forms a binary with a stellar-mass BH, which could dynamically eject other BHs from the cluster and reduce IMBH growth \citep[e.g.,][]{Leigh2014}. Future studies will explore the possibility of modeling the evolution of orbits taking into account GW emission in a more self-consistent manner, with the goal to accurately capture the ratio between true IMRIs and direct plunges.

\section{Acknowledgements} 

We are grateful to I. Andreoni, M. Häberle, F. Rasio, S. Rose, A. Seth, R. Spurzem, and N. Stone for very helpful discussions. We also thank R. van der Marel for sharing his data with us.  
Support for E.G.P.\ was provided by the National Science Foundation Graduate Research Fellowship Program under Grant DGE-2234667.  CR acknowledges support from NASA ATP Grant 80NSSC24K0687, an Alfred P.~Sloan Research Fellowship, and a David and Lucile Packard Foundation Fellowship.  This research was also supported in part through the computational resources and staff contributions provided for the Quest high-performance computing facility at Northwestern University, which is jointly supported by the Office of the Provost, the Office for Research, and Northwestern University Information Technology.  We acknowledge the computing resources provided by North Carolina State University High Performance Computing Services Core Facility.

\bibliographystyle{aasjournal}
\bibliography{main}

\appendix

\section{Loss Cone Dynamics}
\label{sec:rw_method}

\subsection{Prescriptions}
\label{sec:rw_prescripts}

Before the random walk (RW) loop, several key quantities are computed. First, we calculate a representative deflection angle per orbit using Equation \ref{eq:norb}. We also evaluate the tidal radius using Equation \ref{eq:Rdisr} and the Schwarzschild radius for the MBH, adopting the larger of the two as the disruption criterion. 
If a disruption ensues, the accretion fraction will be determined by the choice of the disruption radius. In the case of the tidal radius, $50\%$ of the stellar mass is accreted. If the disruption occurs within  the Schwarzschild radius, then $100\% $ of the mass is accreted.

The length of the RW is set as the total quadratic deflection angle ($L_2 = \delta\theta_{\rm step} ^2 )$, where $\delta\theta_{\rm step}$ is the deflection angle calculated from two-body relaxation in the simulation's timestep. 

During the RW, the magnitude of the object's velocity in the encounter frame ($\omega$) is held constant, while its direction changes randomly. The velocity in the cluster frame is given by $v = w \, + \, v_{CM}$, which is used to evaluate entry into the loss cone. The RW proceeds as follows: 

\begin{enumerate}[label=\arabic*.]

\item The loss cone is calculated using Equation~\ref{eq:LC} with the updated value for $v$. 

\item Next, we test for entry into the loss cone. If $v^{tg} = \sqrt{(v^x_{\rm CM} + w^x)^2 + (v^y_{\rm CM} + w^y)^2 } \leq v_{\rm LC}$, the object has entered the loss cone. We then calculate the time to reach pericenter ($\tau$). If $\tau < t_{\rm step}$, where $t_{\rm step}$ is the time left in the RW step, we proceed as follows:

    \begin{enumerate}[label*=\arabic*.]
    
    \item \textit{Single stars}: If the object was also in the loss cone in the previous step, it is disrupted. Otherwise, we take an additional RW step of the size $\tau$. If the single star remains in the loss cone in the next step and $\tau < t_{\rm step}$ again, it is disrupted.
     
    \item \textit{Binaries}: the encounter is integrated for an orbital period using the \texttt{Fewbody} code. This allows accurate resolution of outcomes, including binary disruption, single or double TDEs, or mergers. 

    \end{enumerate}

\item If the object is not in the loss cone, we evaluate whether it could inspiral due to gravitational wave emission by calculating the inspiral time, $t_{\rm GW}$.  If $t_{\rm GW} < t_{\rm step}$, we assume full accretion for compact objects and $50\%$ accretion for stellar objects.

\item If $L_2 \leq 0 $ the RW ends. If not, the value of $L_{2}$ is updated $\rm L_2 = L_2 -  \Delta^2$, where \mbox{$\Delta = \rm max(\delta\theta_{\rm orb}, min(\Delta_{\rm max},\Delta_{\rm safe}, \sqrt{\rm L_2}))$}, $\Delta_{\rm max} = 0.1\pi, \, \Delta_{\rm safe} = c_{\rm safe}(v^{\rm tg} - v_{\rm LC})/w \, $ with  $c_{\rm safe} = 0.2$. This adaptive step amplitude ensures that the steps get progressively smaller, approaching magnitudes comparable to the orbital period. The only exception is when a binary enters the loss cone and is not disrupted, in which case $\Delta = \delta \theta_{\rm orb}$ to accurately capture the loss cone physics. The new orientation of $\omega$ is set by a random angle with uniform $[0,2\pi]$. The next step begins at point 1. 

\end{enumerate}

\begin{figure}
\begin{center}
\includegraphics[width=0.5\linewidth]{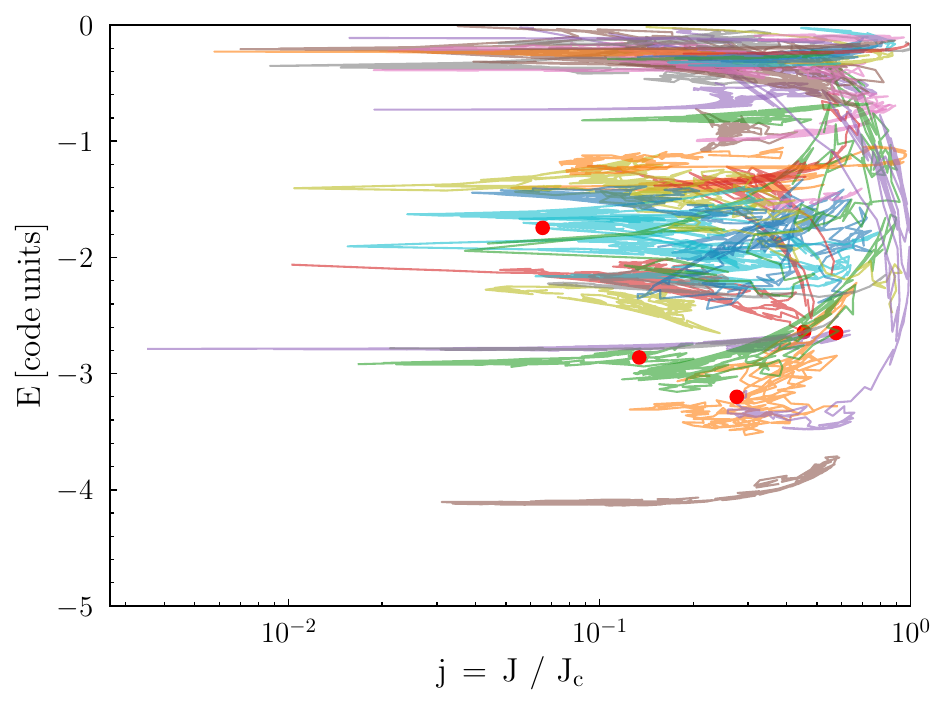}
\caption{\footnotesize \label{fig:RW} Example random walk trajectories from our loss cone treatment, shown in dimensionless specific energy versus dimensionless specific angular momentum (where $\rm J_c$ is the specific angular momentum of an orbit at a given energy). Red dots indicate tidal disruption events. }
\end{center}
\end{figure}

In Figure~\hyperref[fig:RW]{\ref{fig:RW}} we show example RW trajectories in dimensionless specific energy and specific angular momentum space. The red markers indicate tidally disrupted stars. The elongated shapes at low angular momentum (high eccentricities) indicate faster diffusion in angular momentum space than in energy space. In contrast, for high angular momentum orbits (less eccentric orbits), the diffusion is more symmetric. This behavior is consistent with theoretical expectations that diffusion into the loss-cone occurs primarily in angular momentum space\citep[e.g.,][]{Alexander2017}.

\subsection{Comparisons to $N$-Body Models}
\label{sec:comparison}

A valuable benchmark for assessing the accuracy of Monte Carlo approaches is a comparison with direct $N$-body simulations. Particularly relevant to our study are the $N$-body models of single-mass and multi-mass star clusters containing MBHs presented in \citet{Baumgardt2004}. Following previous work \citep[see Figure 5 in][]{Umbreit2012}, we simulated clusters using the same initial conditions as model 16 from \citet{Baumgardt2004}. This single-mass model had an initial number of stars $N \sim 178,800$ (each with mass $1 \, \msun$) initially distributed according to a King profile with concentration parameter $w_0 = 10$, and an initial MBH seed mass of $ 461 \, \msun$. To enable a direct comparison, we disabled both stellar evolution and direct collisions. Furthermore, each star has  a fixed tidal radius of $ 1 \times 10^{-7} $ in $N$-body units. 

\begin{figure}
\begin{center}
\includegraphics[width=\linewidth]{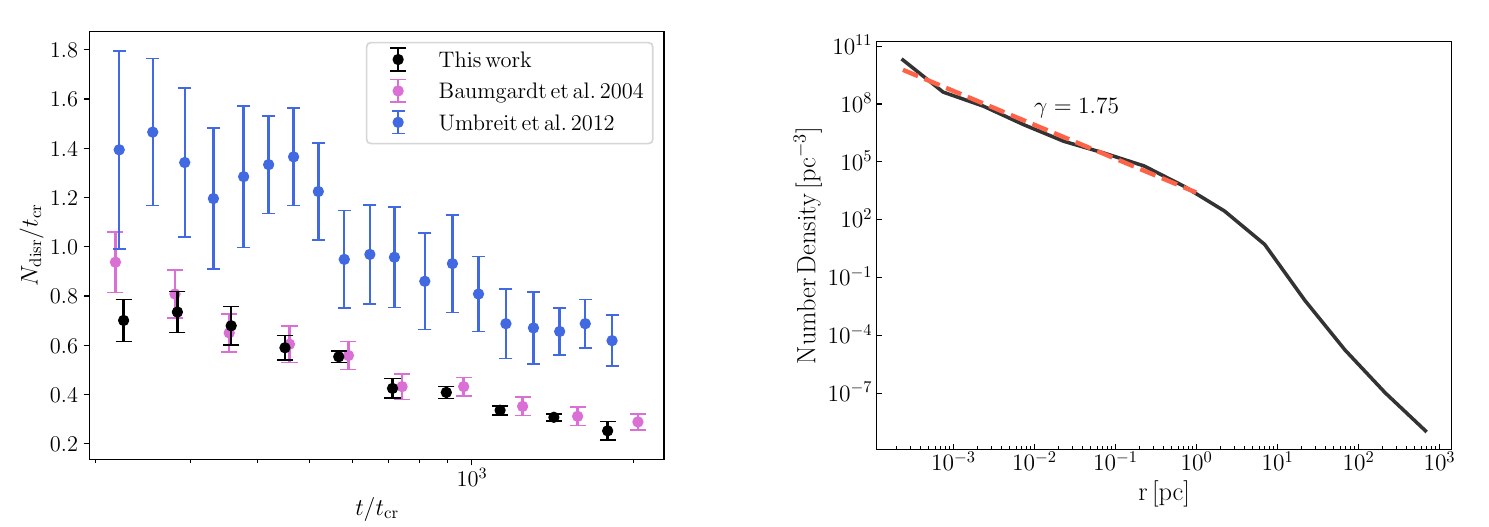}
\caption{\footnotesize \label{fig:Comparisons} \textit{Left}: Comparison of tidal disruption rates per crossing time for our MC simulations (black), previous MC results of \cite{Umbreit2012} (blue), and direct $N$-body model of \cite{Baumgardt2004}(purple). Our models show the average and standard deviation across $5$ realizations of the model, while the previous MC work was averaged over 9 models. The disruption rates and errors of the $N$-body results are time-averages and the standard deviations, respectively. \textit{Right}: Example number density profile for one of the single-mass component models. A density cusp with a power law slope of $\gamma = 1.75 $ forms after $300$ Myr, in agreement with theoretical predictions \citep{Bahcall1976}.}
\end{center}
\end{figure}

On the left panel of Figure~\ref{fig:Comparisons}, we show the mean tidal disruption rate over $2000 $ crossing times, averaged over $5$ realizations, with error bars indicating the standard deviation. For reference, we also include results from direct $N$-body models of \cite{Baumgardt2004} and previous Monte Carlo results of \cite{Umbreit2012}. Our updated loss cone prescriptions show significantly improved agreement with $N$-body simulations. The only noticeable deviation is at early times. This is likely due to the effect of MBH wandering in $N$-body models, which we are not able to capture in our Monte Carlo code. Additionally, since the Monte Carlo method assumes the cluster remains in dynamical equilibrium, it does not accurately capture the initial violent relaxation that occurs when such a MBH is introduced in a stellar population. Thus, some discrepancy with the $N$-body results at early times is expected.

As an additional benchmark, we compute the number density profiles for one of the models at $300$ Myr, shown on the right panel of Figure~\ref{fig:Comparisons}. A central density cusp with a power-law index of $\gamma \approx 1.75$ forms within the MBH's radius of influence, consistent with theoretical predictions for relaxed stellar populations around MBHs \citep{Bahcall1976}.

\subsection{Formation of cusps }
\label{sec:cusps}

Studies of how stellar populations settle around MBHs are crucial to make estimates on the rate of stellar consumptions. \cite{Bahcall1976} showed that a spherically symmetric stellar population of equal-mass point particles with approximately isotropic velocities will settle into a density cusp that can be approximated as $n(r)\propto r^{-\gamma}$, where $\gamma = 7/4$. \cite{Bahcall1977} extended this analysis to a multi-mass stellar distribution and found that a value of $\gamma = 7/4$ holds for the heavy particles, but the lighter ones fall into a more shallow profile with approximately $\gamma = 1.5$. These key theoretical results have been confirmed by N-body simulations \citep{Baumgardt2004,Freitag2006, Baumgardt2019b}. 

\begin{figure}

    \centering
    \includegraphics[width=0.6\linewidth]{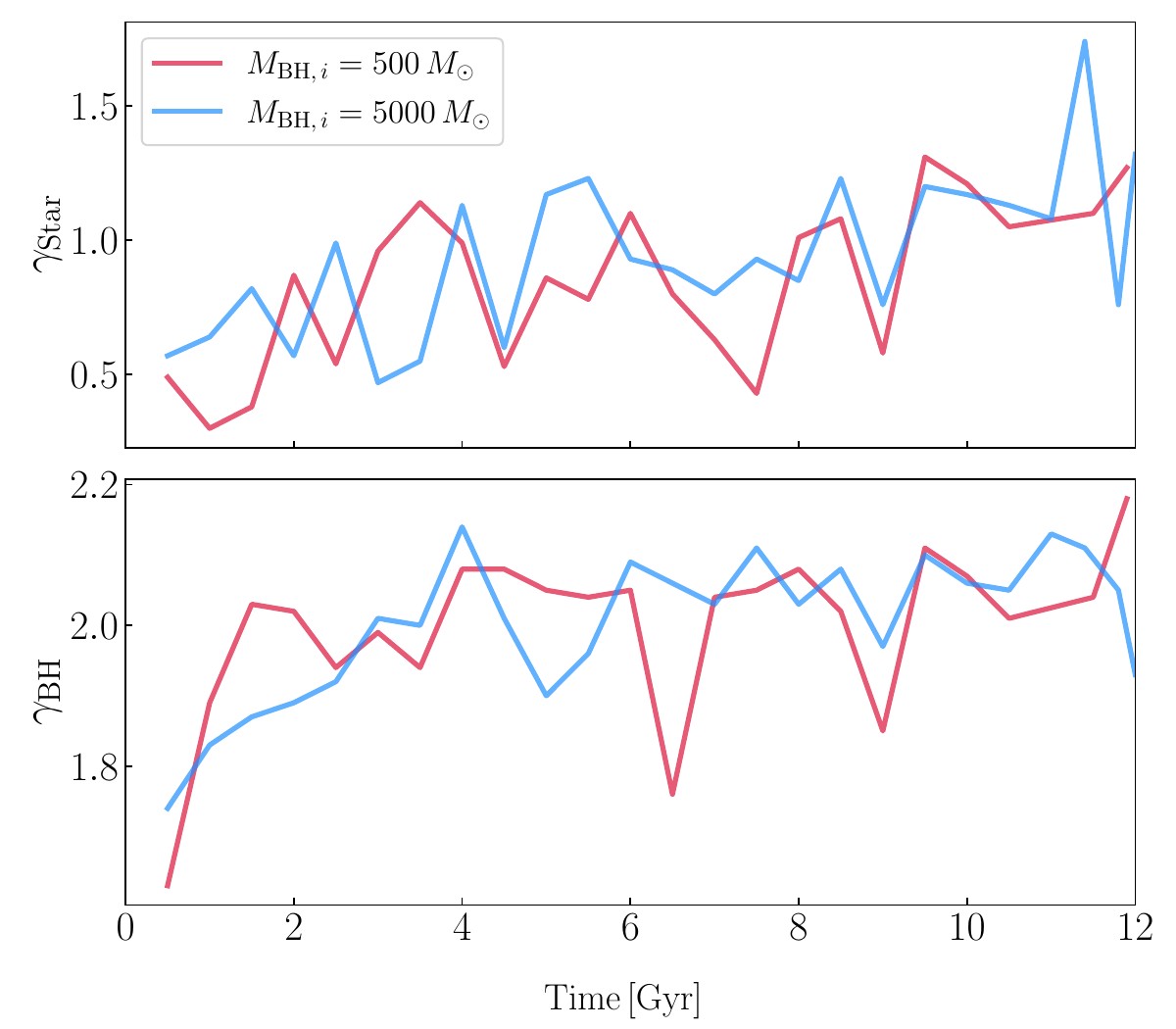}
    \caption{\label{fig:gamma} The time evolution of the mass density power law indices for the stellar (top panel) and BH (bottom panel) populations, fitted in range $[0.001$--$1]$ pc.}
    
\end{figure}

 Here, we compute $3$D mass density profiles at every $0.5$~Gyr and calculate the power-law index for  radial positions between $0.001-1$ pc. In Figure~\ref{fig:gamma} we show the evolution of the power law indices for the stellar and BH populations in our models. We see that the stellar population settles into a cusp with $\gamma \sim 1.3$ in both models, while the BH population exhibits steeper cusps with $\gamma \sim2.1$ and $\sim 1.97$ for the $500 \, \msun$ and $5000 \, \msun$ seed models, respectively. The inferred values are broadly consistent with theoretical predictions, and exhibit variations across epochs. These variations are likely due to the continuous growth of the MBH, different physical processes happening in the cluster such as mass segregation and core expansion, as well as small-number fluctuations in the innermost regions over time. In addition, the steeper cusps we observe for the BHs are likely a result of mass segregation, as \cite{Alexander2009} showed that in the strong mass segregation limit, the heavier components settle into steeper density cusps.

\subsection{King vs. Elson Profiles}
\begin{figure*}
\begin{center}
\includegraphics[width=0.9\linewidth]{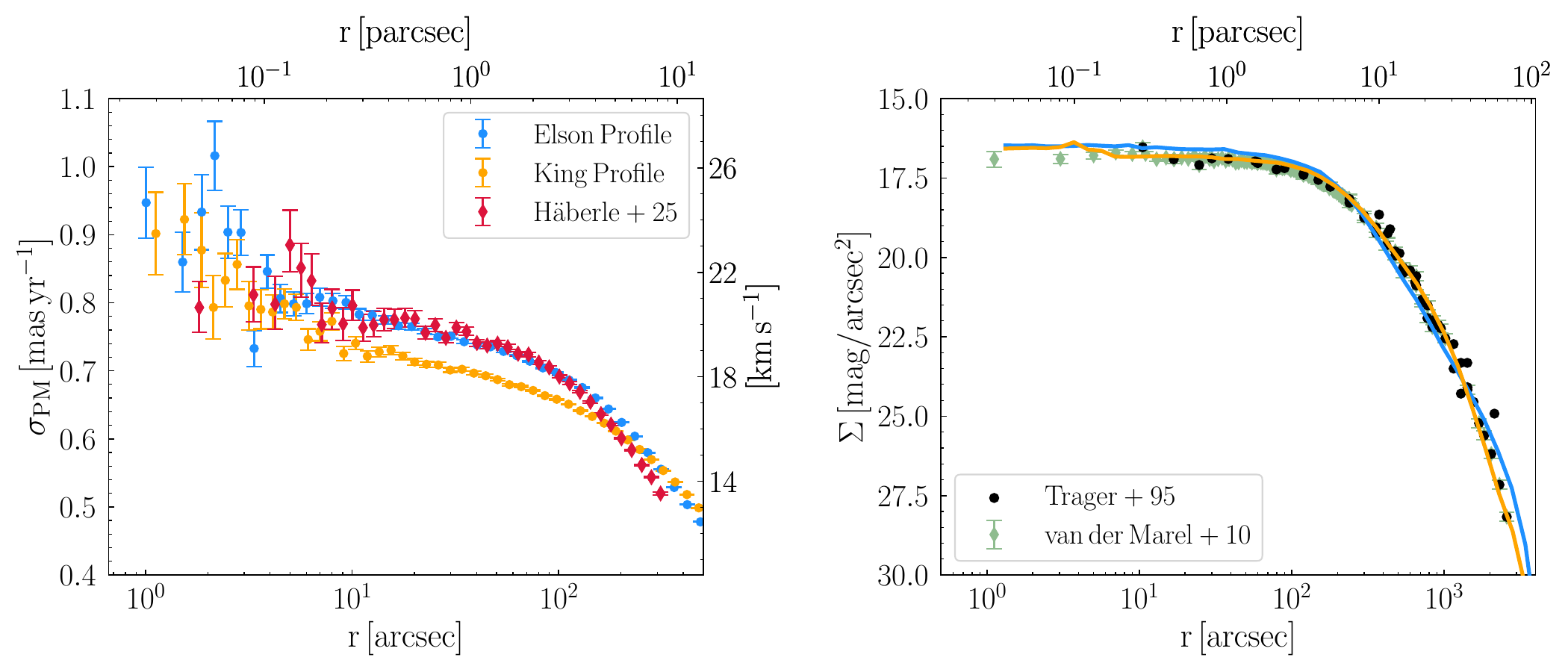}
\caption{\footnotesize \label{fig:ElsonKing} Proper motion velocity dispersion (left) and V-band surface brightness (right) profiles at $12$~Gyr. The Elson profile corresponds to the model presented in the main text with a BH seed of $5000 \, \msun$, while the King model has identical initial conditions but uses a King profile with concentration parameter $w_0 = 8$. Observational data are the same as those shown in Figure~\ref{fig:VDP_SBP}. }
\end{center}

\end{figure*}

Among the broad grid search we conducted to find the models that best match observations of \omegacen, one of the parameters we varied was the initial density distribution. We found that Elson profiles generally yield better-fitting models than King profiles. In Figure~\ref{fig:ElsonKing}, we compare models with identical initial conditions except for their initial density distributions: one follows an Elson profile (the model in the main text), and the other a King profile with a concentration parameter $w_0 = 8 $. In the VDP, the King model significantly deviates from the observed morphology of the cluster. We also explored King profiles with different concentration parameters (not shown), but none provided better agreement. 

\end{document}